\documentclass[aps,prl,twocolumn,groupedaddress]{revtex4}

\usepackage[dvips]{graphicx}
\usepackage{dcolumn}
\usepackage{amsmath,amssymb}
\usepackage{mathrsfs,bm}
\usepackage{pstcol,pstricks,pst-3d,pst-plot,pst-node}
\usepackage{pst-plot,pstricks,pst-node,pst-grad}
\usepackage{epsfig}

\begin{document}


\title{Quantitative and Qualitative Study of Gaussian Beam Visualization Techniques}

\author{J. Magnes}
\email{jenny.magnes@usma.edu}
\homepage{http://www.dean.usma.edu/departments/physics/Research/Magnes.htm}
\author{D. Odera}%
\author{J. Hartke}
\author{M. Fountain}
\author{L. Florence}
\author{V. Davis}
 \affiliation{Department of Physics,
U.S. Military Academy,
West Point, NY 10996-1790.\\
\thanks{ARO}
}%


\date{\today}

\begin{abstract}
 We present a comparative overview of existing laser beam profiling methods.
 We compare the the knife-edge,
scanning slit, and pin-hole methods.  Data is presented in a
comparative fashion.  We also elaborate on the use of CCD profiling
methods and present appropriate imagery.  These methods allow for a
quantitative determination of transverse laser beam-profiles using
inexpensive and accessible methods. The profiling techniques
presented are inexpensive and easily applicable to a variety of
experiments.
\end{abstract}


\maketitle

\section{\label{sec:level1}Introduction}
How large or how "good" a particular laser beam is, can seem like an
infinitely abstract question since a laser beam fades gradually.
However, comparing the effects of an expanded laser beam striking a
piece of paper with burning a piece of paper with the same focused
laser beam brings across the notion that the beam size can make a
difference in many physical scenarios.  Of course, questions on
which methods are appropriate to determine beam quality, divergence
and size arise immediately in these types of discussions.

It is a valuable exercise to agree on some type of quantitative
definition of a laser beam radius. The standards which the
scientific community has agreed upon can then be discussed in a
meaningful way.

\begin{figure}
\begin{center}
\psset{unit=1.0cm}
\begin{pspicture}(-3.2,-0.3)(4,4.8)
\psgrid[subgriddiv=0,griddots=5,gridlabels=0pt](-3,-0.1)(3,4.3)
\psline[linewidth=1.5pt]{->}(-2.5,0)(3,0)
\psline[linewidth=1.5pt]{->}(0,-0.1)(0,4.7)
\psplot[plotstyle=curve,linewidth=1.0pt]{-3}{3}{4 2.7183 x 2 exp exp div}
\pnode(0,4){A} \pnode(-0.9,4){B} \ncline[linestyle=dashed]{A}{B}
\rput*[l](-1.2,4){$I_0$} \pnode(1,1.47){C} \pnode(-1,1.47){D}
\ncline[linestyle=dashed]{C}{D} \rput*[c](-1.7,1.47){$I_0/e$}
\pnode(1.45,.54){E} \pnode(-1.45,.54){F}
\ncline[linestyle=dashed]{E}{F} \rput*[c](-2.25,.54){$I_0/e^2$}
\psline{<->}(0,1.47)(.98,1.47)
\rput*[c](0.44,1.47){$R$}
\psline{<->}(0,.54)(1.4,.54)
\rput*[c](.66,.54){$r$}
\rput[l](2.75,-.25){$x$} \rput[l](-.3,4.5){$I$}
\rput[l](-0.05,-.3){$x_0$}
\end{pspicture}
\caption{\label{fig:Gauss} The Gaussian beam profile of the
irradiance, I, is displayed in this two dimensional figure as a
function of x. The $1/e^2$ beam radius, r, as well as the $1/e$ beam
radius, R, are indicated.}
\end{center}
\end{figure}
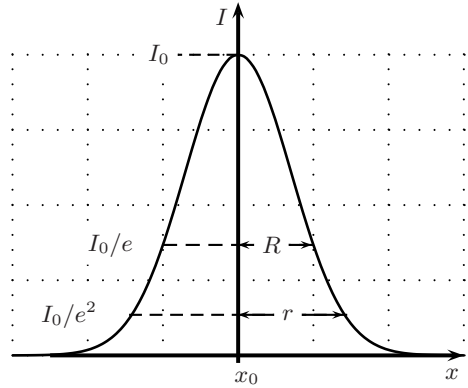

In general, the irradiance, $I(x,y)$, of an ideal laser beam
displays a Gaussian profile as described by Chapple \cite{Slit}:
\begin{equation}
I(x,y)=I_0 \hspace{2pt}
exp\left[-\frac{2((x-x_0)^2+(y-y_0)^2)}{r^2}\right],
\label{eq:Gauss2}
\end{equation}
where $I_{0}$ is the peak irradiance at the center of the beam, $x$
and $y$ are the transverse (cross-sectional) cartesian coordinates
of any point with respect to the center of the beam located at
$(x_0,y_0)$, and r is the $1/e^2$ beam radius. The definition above
assumes a Gaussian distribution for the electric field commonly used
in theory.  When the electric field expression\cite{PP} is squared
we end up with a factor of 2 in the exponent as shown in
 Eq.~(\ref{eq:Gauss2}).  It becomes clear from
Eq.~(\ref{eq:Gauss2}) that at the radius, r, the irradiance drops to
$1/e^2$ of its peak value. The Gaussian distribution of the
irradiance can also be defined as given by McCally \cite{KnifeEdge}:
\begin{equation}
I(x,y)=I_0 \hspace{2pt}
exp\left[-\frac{\left((x-x_0)^2+(y-y_0)^2\right)}{R^2}\right].
\label{eq:Gauss}
\end{equation}
In this case, the beam radius, R, is reached when the irradiance
drops to $1/e$ of its maximum value as shown in
Fig.~\ref{fig:Gauss}. Note that the beam radius, r, in
Eq.~(\ref{eq:Gauss2}) is $\sqrt{2}$ times larger than R.

   We will use the definition given
 by Eq.~(\ref{eq:Gauss2}) for the beam radius, r, in this paper.

We can simply replace the irradiance, I, by the power, P, in
equations~(\ref{eq:Gauss2}) and ~(\ref{eq:Gauss}), since power and
irradiance differ only by a constant factor, i.e., the area.  This
realization comes in handy during measurements.

The measurement of Gaussian laser beams can be challenging
 even though the definitions for the size of Gaussian beam are
 clear.\cite{Power,Focus,NonInv,Inversion}  The challenge in measuring a Gaussian beam profile lies in the nature of
 the laser light and the properties that are to be determined, i.e., whether the
 experimentalist is interested in the size of the beam or the beam structure.
 There are a variety of methods available to measure the properties
 of Gaussian laser beams.  In this paper, we will present a
 convenient overview of commonly used beam-profiling methods based on
 the nature of the laser light and the experimental limitations.
   The data presented here allows for qualitative and quantitative comparison of the
 various beam-profiling methods.

\section{\label{KnifeEdge:level1}Knife-Edge Method}

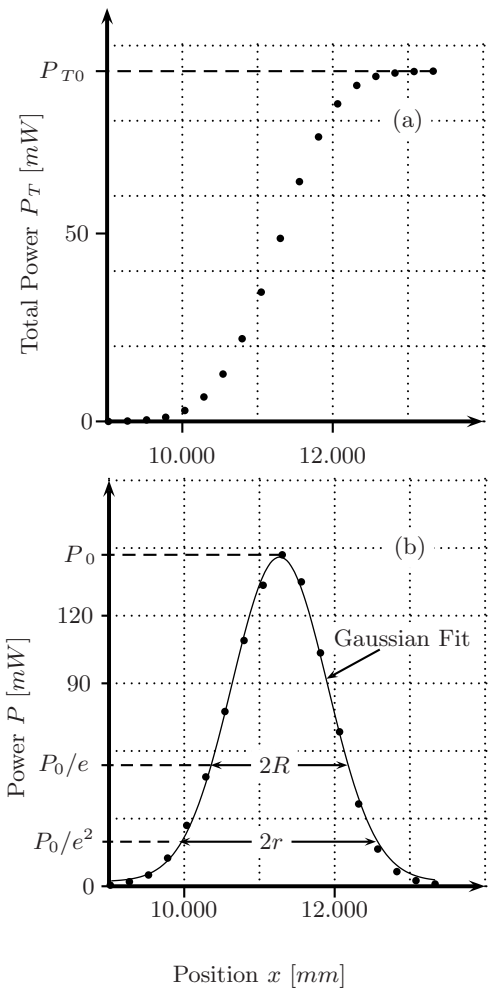
\begin{figure}
\begin{center}
\psset{linecolor=black,xunit=10mm,yunit=0.5mm,runit=1mm}
\begin{pspicture}(12,-15)(10,110)
   \psgrid[xunit=10mm,yunit=10mm,subgriddiv=0,griddots=10,gridlabels=0pt](9,0)(14,5)
    \rput{90}(8,50){Total Power $P\hspace{.5pt}_T$ [$mW$]}
    \pscircle*(13.335,93.2){.5}
    \pscircle*(13.081,93.1){.5}
    \pscircle*(12.827,92.7){.5}
    \pscircle*(12.573,91.8){.5}
    \pscircle*(12.319,89.4){.5}
    \pscircle*(12.065,84.5){.5}
    \pscircle*(11.811,75.7){.5}
    \pscircle*(11.557,63.8){.5}
    \pscircle*(11.303,48.7){.5}
    \pscircle*(11.049,34.4){.5}
    \pscircle*(10.795,22){.5}
    \pscircle*(10.541,12.6){.5}
    \pscircle*(10.287,6.5){.5}
    \pscircle*(10.033,2.9){.5}
    \pscircle*(9.779,1.1){.5}
    \pscircle*(9.525,0.4){.5}
    \pscircle*(9.271,0.1){.5}
    \pscircle*(9.017,0){.5}
    \rput*[c](13,80){(a)}
    \rput[l](8.1,93.2){$P\hspace{1pt}_{T0}$}
   \pnode(8.85,93.2){A} \pnode(13.335,93.2){B}
   \ncline[linestyle=dashed]{A}{B}
   \pnode(10,0){E} \pnode(10,-3){F} \ncline[linestyle=dashed]{E}{F}
\rput[c](10,-9){$10.000$} \pnode(12,0){K} \pnode(12,-3){L}
\ncline[linestyle=dashed]{K}{L} \rput[c](12,-9){$12.000$}
\pnode(9,50){G} \pnode(8.85,50){H} \ncline[linestyle=dashed]{G}{H}
\rput[c](8.6,50){$50$} \pnode(9,0){I} \pnode(8.85,0){J}
\ncline[linestyle=dashed]{I}{J} \rput[l](8.65,0){$0$}
 \psline[linewidth=1.5pt]{->}(9,0)(14,0)
\psline[linewidth=1.5pt]{->}(9,0)(9,110)
\end{pspicture}
\psset{linecolor=black,xunit=10mm,yunit=0.3mm,runit=1mm}
\begin{pspicture}(40,-50)(6.65,180)
    \rput[c]{0}(11,-40){Position $x$ [$mm$]}
    \rput{90}(7.8,75){Power $P$ [$mW$]}
    \psgrid[xunit=10mm,yunit=9mm,subgriddiv=0,griddots=10,gridlabels=0pt](9,0)(14,6)
    \pscircle*(13.335,1){.5}
    \pscircle*(13.081,2.5){.5}
    \pscircle*(12.827,6.5){.5}
    \pscircle*(12.573,16.5){.5}
    \pscircle*(12.319,36.5){.5}
    \pscircle*(12.065,68.5){.5}
    \pscircle*(11.811,103.5){.5}
    \pscircle*(11.557,135){.5}
    \pscircle*(11.303,147){.5}
    \pscircle*(11.049,133.5){.5}
    \pscircle*(10.795,109){.5}
    \pscircle*(10.541,77.5){.5}
    \pscircle*(10.287,48.5){.5}
    \pscircle*(10.033,27){.5}
    \pscircle*(9.779,12.5){.5}
    \pscircle*(9.525,5){.5}
    \pscircle*(9.271,2){.5}
    \pscircle*(9.017,.5){.5}
    \rput*[c](13,150){(b)}
    \rput[l](8.4,147){$P\hspace{1pt}_{0}$}
    \pnode(8.9,147){A} \pnode(11.303,147){B}
    \ncline[linestyle=dashed]{A}{B}
    \pnode(9,90){K} \pnode(8.9,90){L} \ncline[linestyle=dashed]{K}{L}
\rput[c](8.6,90){$90$} \pnode(10,0){E} \pnode(10,-5){F}
\ncline[linestyle=dashed]{E}{F} \rput[c](10,-10){$10.000$}
\pnode(12,0){K} \pnode(12,-5){L} \ncline[linestyle=dashed]{K}{L}
\rput[c](12,-10){$12.000$} \pnode(9,120){G} \pnode(8.9,120){H}
\ncline[linestyle=dashed]{G}{H} \rput[c](8.55,120){$120$}
\pnode(9,0){I} \pnode(8.9,0){J} \ncline[linestyle=dashed]{I}{J}
\rput[l](8.65,0){$0$}
 \psline[linewidth=1.5pt]{->}(9,0)(14,0)
\psline[linewidth=1.5pt]{->}(9,0)(9,180)
\psline{<->}(10.35,53.71)(12.17,53.71) \rput*[c](11.2,53.71){$2R$}
\psline[linestyle=dashed](10.35,53.71)(8.9,53.71)
\rput*[c](8.40,53.71){$P_0/e$} \psline{<->}(9.93,19.76)(12.55,19.76)
\rput*[c](11.16,19.76){$2r$}
\psline[linestyle=dashed](9.8,19.76)(8.9,19.76)
\rput*[c](8.40,19.76){$P_0/e^2$}
\pscurve[showpoints=False,linewidth=0.5pt]{-}%
(9.02992,2.40673)(9.11602,2.5816)(9.20212,2.84705)
(9.28822,3.24205)(9.37432,3.81812)(9.46042,4.64139)(9.54653,5.79415)
(9.63263,7.3752)(9.71873,9.4988)(9.80483,12.29122)(9.89093,15.8848)
(9.97703,20.40889)(10.06314,25.97795)(10.14924,32.67709)
(10.23534,40.54617)(10.32144,49.56393)(10.40754,59.63408)
(10.49364,70.57556)(10.57975,82.11908)(10.66585,93.91171)
(10.75195,105.53047)(10.83805,116.50482)(10.92415,126.34679)
(11.01025,134.58627)(11.09636,140.80789)(11.18246,144.68571)
(11.26856,146.01143)(11.35466,144.71297)(11.44076,140.86094)
(11.52686,134.66227)(11.61297,126.44184)(11.69907,116.61423)
(11.78517,105.64918)(11.78517,105.64918)(11.87127,94.03466)
(11.95737,82.24155)(12.04347,70.69347)(12.12958,59.74417)
(12.21568,49.66383)(12.30178,40.63445)(12.38788,32.75315)
(12.47398,26.04192)(12.56008,20.46144)(12.64619,15.92699)
(12.73229,12.32435)(12.81839,9.52426)(12.90449,7.39435)
(12.99059,5.80825)(13.07669,4.65157)(13.1628,3.8253)(13.2489,3.24702)
(13.335,2.85043)\rput*[c](12.9,110){Gaussian Fit}
\psline{->}(12.6,105)(11.89,92)
\end{pspicture}
\caption{\label{fig:erf} (a)  A Coherent Verdi laser was used for
the collection of this sample data. The total power, $P_T$, passing
by the knife-edge is shown as a function of knife-edge displacement.
(b)  The derivative of the knife-edge data is found by using
Eq.~(\ref{eq:derivative}). The $1/e^2$ radius, r, is $1270\mu m\pm20
\mu m$. }
\end{center}
\end{figure}

The knife-edge method is a beam profiling method that allows for
quick, inexpensive, and accurate determination of the
cross-sectional laser beam parameters.  The knife-edge method
requires a sharp edge (typically a razor blade), a translation stage
with a micrometer and a power meter or an energy meter when working
with pulses. The knife-edge is translated perpendicular to the
direction of propagation of the laser beam. With the knife-edge
initially covering the laser beam, the micrometer can be adjusted in
appropriate increments.  Each data point will show an increase in
total power, $P\hspace{.5pt}_T$, passing by the knife-edge until the
entire beam is detected by the power meter at power
$P\hspace{.5pt}_{T0}$ as shown in Fig.~\ref{fig:erf}. The plot shows
then the two dimensional Gaussian profile integrated over the
displacement of the razor blade, $x$, which is the Error function,
$erf(x)$, describing the total power, $P\hspace{.5pt}_T$, bypassing
the knife edge:
\begin{equation}
P\hspace{.5pt}_T(x)=P_0 r\sqrt{\frac{\pi}{8}}\hspace{2pt}
erf\left[\frac{\sqrt2 \hspace{1pt} x}{r}\right]. \label{eq:erf}
\end{equation}


There are several ways to evaluate the $1/e^2$ radius using this
type of data:
\begin{enumerate}
    \item Fitting the Error function and evaluating fitted parameters in
accordance with Eq.~(\ref{eq:erf}).  This is, however, a more
involved procedure based on the mathematical nature of the error
function, we do not recommend a fit to the error function for
practical reasons and for quick results. We reserve the discussion
of this method for another paper.
    \item Taking the derivative of the data using an algorithm that allows
for smoothing of the derivative is practical.  In this case, the
derivative at any data point was taken by averaging the derivative
of two adjacent data points:
\begin{equation}
\frac{dP_T}{dx}=\frac{1}{2}\left(\frac{y_{i+1}-y_i}
{x_{i+1}-x_i}+\frac{y_i-y_{i-1}}{x_i-x_{i-1}}\right),
\label{eq:derivative}
\end{equation}

gives the power at a particular point $x$ in the beam as shown in
Fig~\ref{fig:erf}b.  This type of derivative may be taken quickly by
using spreadsheets.  The students then have two options of finding
the $1/e^2$ radius, r:
\begin{enumerate}
 \item One can simply estimate by dividing the maximum value in the
derivative plot by $e^2$ and finding the corresponding r value by
linear extrapolation between points. \label{bullet:extrapolate}
 \item Alternatively, a Gaussian can be fit to the data in
Fig~\ref{fig:erf}b using one of the widely available data analysis
programs.  This will also allow for a quick assessment of the
experimental error associated with this fit. We used Origin 7.5 in
this case. It is then instructive to compare these results with the
method described in \ref{bullet:extrapolate}. \label{bullet:fit}
\end{enumerate}
  \end{enumerate}


A few data points are sufficient to determine the size of a Gaussian
laser beam to very high accuracy, limited by the resolution of the
translation stage and the stability of the laser. Using a relatively
inexpensive mechanical micrometer, the resolution can be $2\mu m$ or
better\cite{HighResKnifeEdge}. Diffraction does not have a
significant impact since the light does not spread significantly
after passing the knife-edge.  The knife-edge can therefore fit in
easily at any point an optical set-up while the detector can be
conveniently placed after the knife-edge, depending on the spatial
configuration of the setup.

A drawback of the knife-edge method is the lack of resolved features
in the resulting beam-profile.  Since we are measuring the integral
of the Gaussian beam, features such as hot-spots disappear.  Taking
too many data points can be a disadvantage since any noise will be
amplified by taking the derivative.  Keeping the number of data
points to a minimum will ensure a smoother fit.  Methods allowing
for a more detailed examination of the beam-profile are discussed in
the following sections.

\section{\label{Slit Method:level1}Slit Method}

Using a slit to obtain a beam profile can be achieved by simply
replacing the Knife-Edge in section~\ref{KnifeEdge:level1} by a
narrow slit.  We used a $7 \mu m$ slit to confirm the data we
obtained with the knife-edge.  Since we obtain the same error for
the $1/e^2$ radius, r, for both methods, we see that in this case
most of the error originates from instabilities in the laser beam
independently of the method chosen, i.e., more accurate results can
only be obtained with a more stable laser.

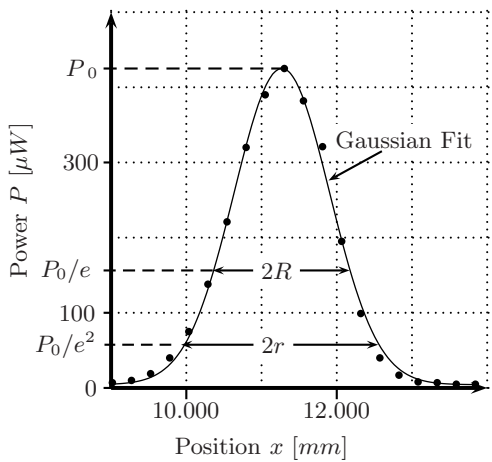
\begin{figure}[h]
\begin{center}
\psset{linecolor=black,xunit=10mm,yunit=0.1mm,runit=1mm}
\begin{pspicture}(12,-90)(10,500)
    \rput[c]{0}(11,-80){Position $x$ [$mm$]}
    \rput{90}(7.8,250){Power $P$ [$\mu W$]}
    \psgrid[xunit=10mm,yunit=10mm,subgriddiv=0,griddots=10,gridlabels=0pt](9,0)(14,5)
    \pscircle*(9.017,7){.5}
    \pscircle*(9.271,10){.5}
    \pscircle*(9.525,19){.5}
    \pscircle*(9.779,40){.5}
    \pscircle*(10.033,75){.5}
    \pscircle*(10.287,138){.5}
    \pscircle*(10.541,221){.5}
    \pscircle*(10.795,320){.5}
    \pscircle*(11.049,390){.5}
    \pscircle*(11.303,425){.5}
    \pscircle*(11.557,382){.5}
    \pscircle*(11.811,321){.5}
    \pscircle*(12.065,195){.5}
    \pscircle*(12.319,99){.5}
    \pscircle*(12.573,40){.5}
    \pscircle*(12.827,17){.5}
    \pscircle*(13.081,8){.5}
    \pscircle*(13.335,7){.5}
    \pscircle*(13.589,5){.5}
    \pscircle*(13.843,5){.5}
    \rput[l](8.4,425){$P\hspace{1pt}_{0}$}
    \pnode(8.9,425){A} \pnode(11.303,425){B}
    \ncline[linestyle=dashed]{A}{B}
    \pnode(9,300){K} \pnode(8.9,300){L} \ncline[linestyle=dashed]{K}{L}
\rput[c](8.6,300){$300$} \pnode(10,0){E} \pnode(10,-15){F}
\ncline[linestyle=dashed]{E}{F} \rput[c](10,-30){$10.000$}
\pnode(12,0){K} \pnode(12,-15){L} \ncline[linestyle=dashed]{K}{L}
\rput[c](12,-30){$12.000$} \pnode(9,100){G} \pnode(8.9,100){H}
\ncline[linestyle=dashed]{G}{H} \rput[c](8.55,100){$100$}
\pnode(9,0){I} \pnode(8.9,0){J} \ncline[linestyle=dashed]{I}{J}
\rput[l](8.65,0){$0$}
 \psline[linewidth=1.5pt]{->}(9,0)(14,0)
\psline[linewidth=1.5pt]{->}(9,0)(9,500)
\psline{<->}(10.35,156.33)(12.17,156.33)
\rput*[c](11.2,156.33){$2R$}
\psline[linestyle=dashed](10.35,156.33)(8.9,156.33)
\rput*[c](8.40,156.33){$P_0/e$}
\psline{<->}(9.93,57.51)(12.55,57.51) \rput*[c](11.16,57.51){$2r$}
\psline[linestyle=dashed](9.8,57.51)(8.9,57.51)
\rput*[c](8.40,57.51){$P_0/e^2$}
\pscurve[showpoints=False,linewidth=0.5pt]{-}%
(9.02131,5.16)(9.10741,5.64)(9.19351,6.38)(9.27961,7.48)(9.36571,9.08)
(9.45181,11.38)(9.53792,14.61)(9.62402,19.06)(9.71012,25.05)
(9.79622,32.95)(9.88232,43.14)(9.96842,56.02)(10.05453,71.91)
(10.14063,91.09)(10.22673,113.69)(10.31283,139.67)
(10.39893,168.78)(10.48503,200.51)(10.57114,234.12)(10.65724,268.6)
(10.74334,302.72)(10.82944,335.12)(10.91554,364.38)(11.00164,389.1)
(11.08775,408.03)(11.17385,420.2)(11.25995,424.95)(11.34605,422.02)
(11.43215,411.56)(11.51825,394.15)(11.60436,370.7)(11.69046,342.4)
(11.77656,310.62)(11.86266,276.77)(11.94876,242.26)
(12.03486,208.35)(12.12097,176.09)(12.20707,146.3)(12.29317,119.55)
(12.37927,96.14)(12.46537,76.15)(12.55147,59.5)(12.63758,45.94)
(12.72368,35.14)(12.80978,26.74)(12.89588,20.33)
(12.98198,15.55)(13.06808,12.06)(13.15419,9.56)(13.24029,7.8)
(13.32639,6.6)(13.41249,5.79)(13.49859,5.26)(13.58469,4.91)
(13.6708,4.69)(13.7569,4.56)(13.843,4.47)
\rput*[c](12.9,330){Gaussian Fit} \psline{->}(12.6,315)(11.89,276)
\end{pspicture}
\caption{\label{fig:slit} The same Coherent Verdi laser used in the
knife-edge measurement was used for the collection of this sample
data. The power, $P$, passing through a $7\mu m$ slit is shown as a
function of slit displacement. The $1/e^2$ radius, r, is found to
agree with the results reported in section~\ref{KnifeEdge:level1}
($1270\mu m\pm20 \mu m$). }
\end{center}
\end{figure}

The slit method assumes that the slit is infinitely narrow so that
we can map the power transmitted through the slit at any point of
the profile.  We can therefore use the slit as long as it is much
narrower than the beam.  The data can then be plotted directly as
shown in Fig.~\ref{fig:slit}.

Traditionally slits have been used as a profiling method when the
slit width is much smaller than the beam itself as it is the case in
our example.  Nevertheless, a slit may also be used when the slit
width is about the same size as the laser beam when measuring the
beam waist, $w_0$, of a focused beam as Chapple\cite{Slit} has
shown:
\begin{equation}
w_0^2=w_{0s}^2-\frac{s^2}{3}, \label{eq:slit}
\end{equation}
where $w_{0s}^2$ is the uncorrected size of the beam waist, and $s$
is the width of the slit.

Even though slits are most appropriate for large beams, care must be
taken that the results are not distorted by the beam exceeding the
longitudinal dimension of the slit.  Diffraction is not an issue as
long as the detector is positioned close enough to the slit to
collect the light after passing through the slit before it spreads
beyond the dimensions of the detector.  A lens can be used right
after the slit to collect the diffracted light if the detector
cannot be placed immediately after the slit.

Finally, the slit method may be impractical when measuring low power
beams since only a very narrow slice of the laser beam passes
through the slit.  This power drop becomes evident when comparing
Figs.~\ref{fig:erf} and~\ref{fig:slit}.

As with the Knife-Edge method, the Slit method does not reveal beam
profile features like hot spots.  However, these methods are most
practical for determining the size of a Gaussian laser beam.

\section{\label{Pinhole Method:level1}Pinhole Method}

The pinhole method allows the user to scan across a laser beam and
record the irradiance at every point.  This technique allows for the
collection of a highly resolved beam profile.  Hot spots and other
beam features can then be detected.  Intensity distributions across
a laser beam become critical for experiments involving optical
non-linearities.


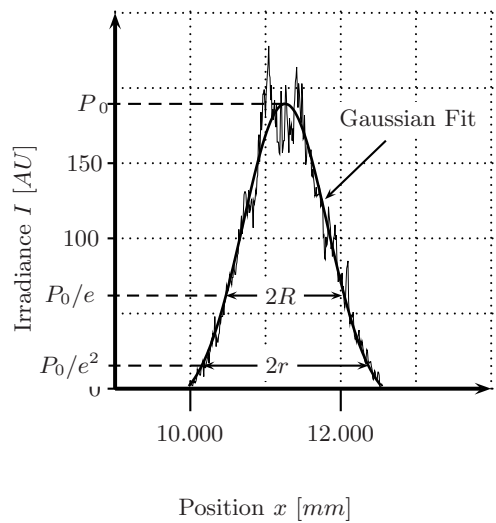
\begin{figure}[h]
\begin{center}
\psset{linecolor=black,xunit=10mm,yunit=0.2mm,runit=1mm}
\begin{pspicture}(12,-90)(10,250)
    \rput[c]{0}(11,-80){Position $x$ [$mm$]}
    \rput{90}(7.8,100){Irradiance $I$ [$AU$]}
    \psgrid[xunit=10mm,yunit=10mm,subgriddiv=0,griddots=10,gridlabels=0pt](9,0)(14,5)
    \rput[l](8.5,189.4651){$P\hspace{1pt}_{0}$}
    \pnode(8.9,189.4651){A} \pnode(11.25537,189.4651){B}
    \ncline[linestyle=dashed]{A}{B}
    \pnode(9,150){K} \pnode(8.9,150){L} \ncline[linestyle=dashed]{K}{L}
\rput[c](8.6,150){$150$} \pnode(10,0){E} \pnode(10,-15){F}
\ncline[linestyle=dashed]{E}{F} \rput[c](10,-30){$10.000$}
\pnode(12,0){K} \pnode(12,-15){L} \ncline[linestyle=dashed]{K}{L}
\rput[c](12,-30){$12.000$} \pnode(9,100){G} \pnode(8.9,100){H}
\ncline[linestyle=dashed]{G}{H} \rput[c](8.55,100){$100$}
\pnode(9,0){I} \pnode(8.9,0){J} \ncline[linestyle=dashed]{I}{J}
\rput[l](8.65,0){$0$}
 \psline[linewidth=1.5pt]{->}(9,0)(14,0)
\psline[linewidth=1.5pt]{->}(9,0)(9,250)
\psline{<->}(10.47,61.94)(12.02,61.94) \rput*[c](11.2,61.94){$2R$}
\psline[linestyle=dashed](10.35,61.94)(8.9,61.94)
\rput*[c](8.40,61.94){$P_0/e$}
\psline{<->}(10.17,15.33)(12.37,15.33) \rput*[c](11.16,15.33){$2r$}
\psline[linestyle=dashed](10.17,15.33)(8.9,15.33)
\rput*[c](8.40,15.33){$P_0/e^2$}
\psline[linewidth=.1]%
(9.98023,3)(9.98886,3)(9.9975,5)(10.00613,5)(10.01477,3)(10.0234,2)
(10.03204,2)(10.04068,5)(10.04931,4)(10.05795,5)(10.06658,4)
(10.07522,5)(10.08385,7)(10.09249,13)(10.10112,11)(10.10976,12)
(10.1184,10)(10.12703,9)(10.13567,9)(10.1443,14)(10.15294,15)
(10.16157,18)(10.17021,15)(10.17885,8)(10.18748,20)(10.19612,14)
(10.20475,14)(10.21339,17)(10.22202,19)(10.23066,21)(10.23929,26)
(10.24793,17)(10.25657,24)(10.2652,25)(10.27384,31)(10.28247,33)
(10.29111,30)(10.29974,37)(10.30838,35)(10.31702,37)(10.32565,48)
(10.33429,38)(10.34292,28)(10.35156,34)(10.36019,31)(10.36883,31)
(10.37747,36)(10.3861,37)(10.39474,46)(10.40337,44)(10.41201,48)
(10.42064,48)(10.42928,48)(10.43791,54)(10.44655,49)(10.45519,57)
(10.46382,58)(10.47246,62)(10.48109,69)(10.48973,67)(10.49836,65)
(10.507,66)(10.51564,73)(10.52427,76)(10.53291,73)(10.54154,71)
(10.55018,69)(10.55881,74)(10.56745,74)(10.57608,85)(10.58472,79)
(10.59336,76)(10.60199,76)(10.61063,82)(10.61926,84)(10.6279,89)
(10.63653,90)(10.64517,106)(10.65381,98)(10.66244,99)(10.67108,109)
(10.67971,115)(10.68835,103)(10.69698,108)(10.70562,118)(10.71426,129)
(10.72289,131)(10.73153,131)(10.74016,129)(10.7488,118)(10.75743,115)
(10.76607,120)(10.7747,132)(10.78334,117)(10.79198,121)(10.80061,119)
(10.80925,125)(10.81788,126)(10.82652,120)(10.83515,107)(10.84379,125)
(10.85243,123)(10.86106,126)(10.8697,132)(10.87833,130)(10.88697,140)
(10.8956,142)(10.90424,165)(10.91287,169)(10.92151,151)(10.93015,165)
(10.93878,184)(10.94742,172)(10.95605,179)(10.96469,196)(10.97332,196)
(10.98196,199)(10.9906,190)(10.99923,181)(11.00787,196)(11.0165,203)
(11.02514,210)(11.03377,216)(11.04241,228)(11.05104,209)(11.05968,207)
(11.06832,192)(11.07695,202)(11.08559,180)(11.09422,186)(11.10286,179)
(11.11149,187)(11.12013,192)(11.12877,177)(11.1374,174)(11.14604,189)
(11.15467,164)(11.16331,164)(11.17194,164)(11.18058,157)(11.18922,156)
(11.19785,182)(11.20649,198)(11.21512,186)(11.22376,157)(11.23239,160)
(11.24103,165)(11.24966,165)(11.2583,183)(11.26694,202)(11.27557,197)
(11.28421,183)(11.29284,176)(11.30148,170)(11.31011,157)(11.31875,180)
(11.32739,169)(11.33602,174)(11.34466,176)(11.35329,165)(11.36193,173)
(11.37056,163)(11.3792,187)(11.38783,192)(11.39647,204)(11.40511,207)
(11.41374,214)(11.42238,200)(11.43101,187)(11.43965,203)(11.44828,198)
(11.45692,190)(11.46556,181)(11.47419,177)(11.48283,203)(11.49146,194)
(11.5001,166)(11.50873,167)(11.51737,182)(11.52601,165)(11.53464,149)
(11.54328,152)(11.55191,161)(11.56055,161)(11.56918,160)(11.57782,157)
(11.58645,155)(11.59509,154)(11.60373,159)(11.61236,142)(11.621,141)
(11.62963,146)(11.63827,140)(11.6469,136)(11.65554,128)(11.66418,129)
(11.67281,134)(11.68145,128)(11.69008,129)(11.69872,129)(11.70735,129)
(11.71599,135)(11.72462,131)(11.73326,121)(11.7419,122)(11.75053,126)
(11.75917,125)(11.7678,101)(11.77644,102)(11.78507,87)(11.79371,86)
(11.80235,101)(11.81098,95)(11.81962,96)(11.82825,102)(11.83689,121)
(11.84552,115)(11.85416,111)(11.8628,107)(11.87143,93)(11.88007,97)
(11.8887,88)(11.89734,93)(11.90597,100)(11.91461,102)(11.92324,109)
(11.93188,99)(11.94052,102)(11.94915,85)(11.95779,77)(11.96642,71)
(11.97506,76)(11.98369,73)(11.99233,73)(12.00097,75)(12.0096,72)
(12.01824,77)(12.02687,64)(12.03551,68)(12.04414,58)(12.05278,54)
(12.06141,58)(12.07005,70)(12.07869,80)(12.08732,82)(12.09596,85)
(12.10459,58)(12.11323,49)(12.12186,43)(12.1305,46)(12.13914,48)
(12.14777,49)(12.15641,48)(12.16504,51)(12.17368,41)(12.18231,35)
(12.19095,33)(12.19959,34)(12.20822,31)(12.21686,34)(12.22549,34)
(12.23413,36)(12.24276,39)(12.2514,31)(12.26003,29)(12.26867,32)
(12.27731,29)(12.28594,31)(12.29458,27)(12.30321,25)(12.31185,20)
(12.32048,18)(12.32912,17)(12.33776,21)(12.34639,26)(12.35503,25)
(12.36366,19)(12.3723,17)(12.38093,17)(12.38957,18)(12.3982,18)
(12.40684,15)(12.41548,13)(12.42411,7)(12.43275,6)(12.44138,8)
(12.45002,12)(12.45865,11)(12.46729,7)(12.47593,6)(12.48456,6)
(12.4932,11)(12.50183,11)(12.51047,9)(12.5191,1)(12.52774,0)
(12.53638,2)(12.54501,3)(12.55365,2)(12.56228,2)
\pscurve[showpoints=False,linewidth=1pt]{-}%
(9.98023,1.67667)(10.03336,5.15135)(10.08649,9.25329)
(10.13962,14.04261)(10.19275,19.57215)(10.24588,25.88376)
(10.29901,33.00444)(10.35214,40.94237)(10.40527,49.68312)
(10.45841,59.18638)(10.51154,69.38327)(10.56467,80.17471)
(10.6178,91.43102)(10.67093,102.99279)(10.72406,114.67351)
(10.77719,126.26353)(10.83032,137.53579)(10.88346,148.25268)
(10.93659,158.17419)(10.98972,167.06662)(11.04285,174.7117)
(11.09598,180.9154)(11.14911,185.51616)(11.20224,188.39185)
(11.25537,189.46515)(11.3085,188.707)(11.36164,186.13794)
(11.41477,181.82715)(11.4679,175.88937)(11.52103,168.47983)
(11.57416,159.7875)(11.62729,150.0271)(11.68042,139.43031)
(11.73355,128.2367)(11.78668,116.68488)(11.83982,105.00429)
(11.89295,93.40798)(11.94608,82.08683)(11.99921,71.20511)
(12.05234,60.89776)(12.10547,51.26915)(12.1586,42.39325)
(12.21173,34.31516)(12.26486,27.05354)(12.318,20.60385)
(12.37113,14.94206)(12.42426,10.0286)(12.47739,5.81225)
(12.53052,2.23389)
\rput*[c](12.9,180){Gaussian Fit} \psline{->}(12.6,165)(11.79,126)
\end{pspicture}
\caption{\label{fig:pinhole} The same Coherent Verdi laser beam was
probed for the collection of this sample data. Difficulty in
locating beam center and the exact location of the pinhole inside
the BeamProfiler by Photon Inc., contributed to the systematic error
which resulted in a measured beam width of $1110\mu m\pm10 \mu m$
deviating from the results of sections \ref{KnifeEdge:level1} and
~\ref{Slit Method:level1}.}
\end{center}
\end{figure}

The beam profile in Fig.~\ref{fig:pinhole} was obtained using an
computer controlled pinhole by Photon Inc.  It is impractical to
mount a pinhole on a translation station and take the data manually
as we did when we used the knife-edge or slit.  It is more
advantageous to use the pinhole in an automated system that allows
for quicker data collection.  In our case, the pinhole and sensor
are contained in one unit so that the exact position of the pinhole
and sensor could not be located easily.  Also, there is no insight
into the mechanism of determining the beam center for the horizontal
sweep of the pinhole resulting in another possible source of
systematic error. These two systematic errors account for a beam
width of $1110\mu m\pm10 \mu m$ deviating from the results in
sections \ref{KnifeEdge:level1} and ~\ref{Slit Method:level1} by
$160\mu m$.

The light passing through the pinhole is significantly reduced in
intensity compared to the slit and therefore even harder to detect
and more prone to poor signal to noise ratio issues.  It should be
noted that it can be difficult to sweep exactly across the center of
the laser beam.  The beam can thus appear smaller than it really is.

\section{\label{CCD:level1}CCD Cameras}

\begin{figure}[h]
\includegraphics[width=8.5cm]{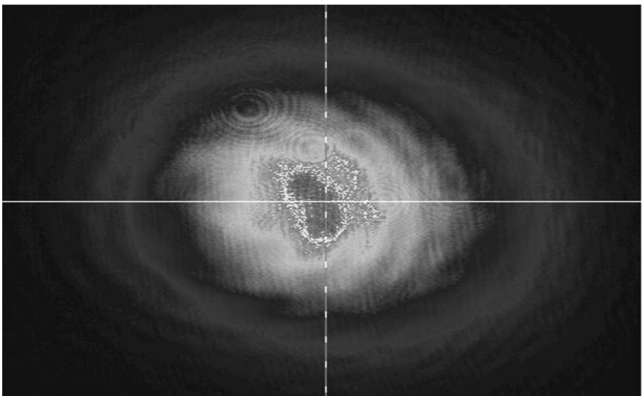}
\caption{\label{fig:CCD} CCD image of CW Verdi laser beam.  The
image reveals a somewhat elliptical shape that shows a combination
of modes. Artifacts from due to interference are visible.}
\end{figure}

Quick qualitative analysis of transverse laser beam profiles can be
achieved through the use of CCD cameras as shown in
Fig.~\ref{fig:CCD}.
 Visual inspection of this beam profile reveals not only an almost
 elliptical shape but also other irregularities in the profile.

There are many different types of CCD cameras available.  The
resolution of a CCD camera is governed by pixel-size and number of
pixels. An adjustable dark current can prevent the CCD camera from
saturating without the use of attenuators.  An inexpensive digital
camera can also be used to capture a beam profile as long as the
beam is carefully attenuated to prevent saturation or damage to the
camera. It should be noted that attenuation tends to distort the
beam profile and beam size.  High-intensity or pulsed lasers may
introduce non-linearities in experimental camera optics leading to a
distorted beam profile. Interference effects originating from dust
and the use of a neutral density filter are visible in
Fig.~\ref{fig:CCD} and should not be confused with beam features.

\section{\label{misc:level1}Creating a Beam Map}

\begin{figure}[h]
\begin{center}
\psset{linecolor=black,xunit=10mm,yunit=0.1mm}
\begin{pspicture}(-6,100)(4.5,500)
   \psgrid[xunit=10mm,yunit=10mm,subgriddiv=1,griddots=10,gridlabels=0pt](-4,2)(2,5)
    \rput{90}(-5.5,350){Radius $r$ [$\mu m$]}
    \pscurve[showpoints=true,linewidth=0.5pt]{-}
    (-4,440)(-2,375)(0,342)(2,352)
    \pscurve[showpoints=true,linestyle=dashed,linewidth=1.2pt,dotsize=.1]{-}%
    (-4,307)(-2,303)(0,302)(2,311)
\psline(0,362)(0,322)\psline(2,372)(2,332)\psline(-2,335)(-2,415)\psline(-4,480)(-4,400)
\psline(-.1,362)(.1,362)\psline(1.9,372)(2.1,372)
\psline(-2.1,335)(-1.9,335)\psline(-3.9,480)(-4.1,480)
\psline(-.1,322)(.1,322)\psline(1.9,332)(2.1,332)
\psline(-2.1,415)(-1.9,415)\psline(-3.9,400)(-4.1,400)
\psline(-4,298)(-4,316)\psline(-2,293)(-2,313)\psline(0,294)(0,310)\psline(2,301)(2,321)
\psline(-3.9,298)(-4.1,298)\psline(-1.9,293)(-2.1,293)
\psline(-.1,294)(0.1,294)\psline(1.9,301)(2.1,301)
\psline(-3.9,316)(-4.1,316)\psline(-1.9,313)(-2.1,313)
\psline(-.1,310)(0.1,310)\psline(1.9,321)(2.1,321)
    \rput*[c](-1,450){(a) Side View of Beam Divergence}
    \pnode(-4,200){C} \pnode(-4,180){D}
\ncline[linestyle=dashed]{C}{D} \rput[c](-4,170){$-4.00$}
\pnode(-2,200){E} \pnode(-2,180){F} \ncline[linestyle=dashed]{E}{F}
\rput[c](-2,170){$-2.00$} \pnode(0,200){G} \pnode(0,180){H}
\ncline[linestyle=dashed]{G}{H} \rput[c](0,170){$0$}
\pnode(-4,200){I} \pnode(-4.2,200){J}
\ncline[linestyle=dashed]{I}{J} \rput[l](-4.8,200){$200$}
\pnode(-4,400){K} \pnode(-4.2,400){L}
\ncline[linestyle=dashed]{K}{L} \rput[l](-4.8,400){$400$}
 \psline[linewidth=1.5pt]{->}(-4,200)(2,200)
\psline[linewidth=1.5pt]{->}(-4,200)(-4,500)
\end{pspicture}
\psset{linecolor=black,xunit=10mm,yunit=30mm}
\begin{pspicture}(-6,0)(7,1.5)
   \psgrid[xunit=10mm,yunit=15mm,subgriddiv=1,griddots=10,gridlabels=0pt](-4,1)(2,3)
    \rput{90}(-5.2,1){Position $y$ [$mm$]}
    \rput{90}(-5.7,.865){Horizontal}
    \rput(-1,0.2){Position $z$ [$mm$]}
    \psline[showpoints=true,linestyle=dashed,linewidth=1.2pt,dotsize=.1]{-}%
    (-4,.815)(2,.7975)
        \psdots(0,.7919)(-2,.81)
    \psline[showpoints=true,linewidth=0.5pt]{-}%
    (-4,.6204)(2,.7174)
        \psdots(-2,.6436)(0,.7)
    \rput*[c](-1,1.3){(b)  Top View of Beam Propagation}
    \pnode(-4,.5){C} \pnode(-4,.45){D}
\ncline[linestyle=dashed]{C}{D} \rput[c](-4,.4){$-4.00$}
\pnode(-2,.5){E} \pnode(-2,.45){F} \ncline[linestyle=dashed]{E}{F}
\rput[c](-2,.4){$-2.00$} \pnode(0,0.5){G} \pnode(0,.45){H}
\ncline[linestyle=dashed]{G}{H} \rput[c](0,.4){$0$} \pnode(-4,1){K}
\pnode(-4.2,1){L} \ncline[linestyle=dashed]{K}{L}
\rput[l](-4.8,1){$1.00$} \pnode(-4,.5){M} \pnode(-4.2,.5){N}
\ncline[linestyle=dashed]{M}{N} \rput[l](-4.8,.5){$0.50$}
\psline[linewidth=1.5pt]{->}(-4,.5)(2,.5)
\psline[linewidth=1.5pt]{->}(-4,.5)(-4,1.5)
\end{pspicture}
\caption{\label{fig:map} A map of two pulsed laser beams is created
to verify beam overlap around experimental focal points.  The laser
used is a Continuum Leopard picosecond system.  (a) A side view of
the beam divergence is presented.   (b) The center positions of the
beams are mapped in a top view to evaluate beam overlap. The laser
beam shown in the solid line needs to be adjusted horizontally to
ensure overlap at $z = 0 mm$.  The error is around $1\%$ for each
data point and too small to show on this scale.}
\end{center}
\end{figure}
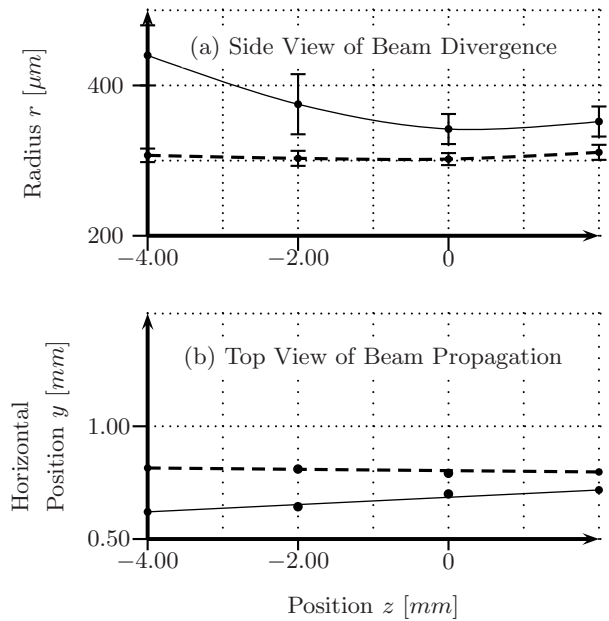

The knife-edge method can even be used to determine beam overlap and
relative beam propagation when using several beams.  Also, beam
divergence can be determined by measuring the beam radius at various
points along the beam path.  The type of beam propagation data shown
in Fig.~\ref{fig:map} is useful in many research projects or may
simply be used to demonstrate Gaussian beam propagation in an
undergraduate laboratory.\cite{PP}

\section{\label{Conclusion:level1}Conclusion}

We have demonstrated the reliability of the knife-edge and slit
beam-profiling methods in determining the size of a Gaussian laser
beam by presenting the identical results of $1270\mu m\pm20 \mu m$
for the beam radius for both methods whereas the pinhole method
shows deviations as presented above. Furthermore, we have presented
two beam profiling methods, the CCD array and the pinhole method, in
evaluating beam quality. There are many sophisticated beam profiling
instruments that are commercially available. Nevertheless, beam
profiles can be obtained effectively with modest resources. Most of
the commercially available beam profilers are based on the
principles outlined in this paper. Under certain circumstances, it
can certainly be convenient and a time saver to purchase a beam
profiling system.

\begin{acknowledgments}
We wish to acknowledge the support of the Army Research Office.  We
appreciate discussions with Dr. Noureddine Melikechi from the
Applied Optics Center at Delaware State University in Dover,
Delaware.
\end{acknowledgments}
\bibliography{beampro}

\begin{thebibliography}{8}
\expandafter\ifx\csname natexlab\endcsname\relax\def\natexlab#1{#1}\fi
\expandafter\ifx\csname bibnamefont\endcsname\relax
  \def\bibnamefont#1{#1}\fi
\expandafter\ifx\csname bibfnamefont\endcsname\relax
  \def\bibfnamefont#1{#1}\fi
\expandafter\ifx\csname citenamefont\endcsname\relax
  \def\citenamefont#1{#1}\fi
\expandafter\ifx\csname url\endcsname\relax
  \def\url#1{\texttt{#1}}\fi
\expandafter\ifx\csname urlprefix\endcsname\relax\def\urlprefix{URL }\fi
\providecommand{\bibinfo}[2]{#2}
\providecommand{\eprint}[2][]{\url{#2}}

\bibitem[{\citenamefont{Chapple}(1994)}]{Slit}
\bibinfo{author}{\bibfnamefont{P.~B.} \bibnamefont{Chapple}},
  \bibinfo{journal}{Opt.\ Eng.} \textbf{\bibinfo{volume}{33}},
  \bibinfo{pages}{2461} (\bibinfo{year}{1994}).

\bibitem[{\citenamefont{Pedrotti and Pedrotti}(1993)}]{PP}
\bibinfo{author}{\bibfnamefont{F.~L.} \bibnamefont{Pedrotti}} \bibnamefont{and}
  \bibinfo{author}{\bibfnamefont{L.~S.} \bibnamefont{Pedrotti}},
  \emph{\bibinfo{title}{Introduction to Optics}} (\bibinfo{publisher}{Prentice
  Hall, New Jersey}, \bibinfo{year}{1993}), \bibinfo{edition}{2nd} ed.

\bibitem[{\citenamefont{McCally}(1984)}]{KnifeEdge}
\bibinfo{author}{\bibfnamefont{R.~L.} \bibnamefont{McCally}},
  \bibinfo{journal}{Appl.\ Opt.} \textbf{\bibinfo{volume}{23}},
  \bibinfo{pages}{2227} (\bibinfo{year}{1984}).

\bibitem[{\citenamefont{Riza and Mughal}(2004)}]{Power}
\bibinfo{author}{\bibfnamefont{A.~R.} \bibnamefont{Riza}} \bibnamefont{and}
  \bibinfo{author}{\bibfnamefont{M.~J.} \bibnamefont{Mughal}},
  \bibinfo{journal}{Opt.\ Eng.} \textbf{\bibinfo{volume}{43}},
  \bibinfo{pages}{793} (\bibinfo{year}{2004}).

\bibitem[{\citenamefont{Skinner and Whitcher}(1971)}]{Focus}
\bibinfo{author}{\bibfnamefont{D.~R.} \bibnamefont{Skinner}} \bibnamefont{and}
  \bibinfo{author}{\bibfnamefont{R.~E.} \bibnamefont{Whitcher}},
  \bibinfo{journal}{J. Phys. E} \textbf{\bibinfo{volume}{5}},
  \bibinfo{pages}{237} (\bibinfo{year}{1971}).

\bibitem[{\citenamefont{Riza and Jorgesen}(2004)}]{NonInv}
\bibinfo{author}{\bibfnamefont{N.~A.} \bibnamefont{Riza}} \bibnamefont{and}
  \bibinfo{author}{\bibfnamefont{D.}~\bibnamefont{Jorgesen}},
  \bibinfo{journal}{Opt. Express} \textbf{\bibinfo{volume}{12}},
  \bibinfo{pages}{1892} (\bibinfo{year}{2004}).

\bibitem[{\citenamefont{Khosrofian and Garetz}(1983)}]{Inversion}
\bibinfo{author}{\bibfnamefont{J.~M.} \bibnamefont{Khosrofian}}
  \bibnamefont{and} \bibinfo{author}{\bibfnamefont{B.~A.}
  \bibnamefont{Garetz}}, \bibinfo{journal}{Appl. Opt.}
  \textbf{\bibinfo{volume}{22}}, \bibinfo{pages}{3406} (\bibinfo{year}{1983}).

\bibitem[{\citenamefont{Plass et~al.}(1997)\citenamefont{Plass, Maestle,
  Wittig, Voss, and Giesen}}]{HighResKnifeEdge}
\bibinfo{author}{\bibfnamefont{W.}~\bibnamefont{Plass}},
  \bibinfo{author}{\bibfnamefont{R.}~\bibnamefont{Maestle}},
  \bibinfo{author}{\bibfnamefont{K.}~\bibnamefont{Wittig}},
  \bibinfo{author}{\bibfnamefont{A.}~\bibnamefont{Voss}}, \bibnamefont{and}
  \bibinfo{author}{\bibfnamefont{A.}~\bibnamefont{Giesen}},
  \bibinfo{journal}{Opt.\ Com.} \textbf{\bibinfo{volume}{134}},
  \bibinfo{pages}{21} (\bibinfo{year}{1997}).

\end{thebibliography}

\end{document}